\documentstyle[aps,prl,floats,epsf,amsfonts]{revtex}

\begin{document}
\draft

\twocolumn[\hsize\textwidth\columnwidth\hsize\csname@twocolumnfalse\endcsname
\title{Fluctuations and correlations in population models with age structure}
\author{Martin Howard$^1$ and R.K.P. Zia$^2$}
\address{$^1$ Department of Physics, Simon Fraser University, Burnaby,
British Columbia, Canada V5A 1S6}
\address{$^2$ Physics Department, Virginia Polytechnic Institute and
State University, Blacksburg, Virginia 24061-0435}  
\date{\today}
\maketitle
\begin{abstract}
We study the population profile in a simple discrete time model of
population dynamics. Our model, which is closely related to certain
``bit--string'' models of evolution, incorporates competition for
resources via a population dependent death probability, as well as a
variable reproduction probability for each individual as a function of
age. We first solve for the steady--state of the model in mean field
theory, before developing analytic techniques to compute Gaussian
fluctuation corrections around the mean field fixed point. Our
computations are found to be in good agreement with Monte--Carlo
simulations. Finally we discuss how similar methods may be applied to
fluctuations in continuous time population models.  
\end{abstract}

]

The problem of population dynamics has attracted enormous interest over many
years (for some introductions and recent applications see Refs.~\cite
{murray,pielou,pollard,charles,rose,penna,tol,redner}). Beginning with simple
logistic growth models \cite{murray}, a tremendous variety of systems have
been studied displaying a diverse range of behavior, varying from stable
fixed points to strange attractors. Broadly speaking these models split
naturally into two categories: those using continuous and those using
discrete time. The simplest discrete time models describe species where
there is no overlap between successive generations, leading to difference
equations of the form $N(t+1)=g[N(t)]$, where $N(t)$ is the total population
at time $t$. However these models may easily be generalized to species with
multiple discrete age generations (for example: eggs, larvae, adults), where
one or more generations may be present simultaneously. Instead of a single
variable $N$, information about the age distribution is now carried in a
``vector'' ${\bf n}(t)\equiv \{n_0(t),n_1(t),\ldots ,n_D(t)\}$, where $%
n_a(t) $ is the number of individuals of age $a$ at time $t$. Note that $D$
is the maximum age, while $n_0$ stands for the number of ``newborns''.
Beginning with the pioneering work of Leslie \cite{leslie} models of this
type have been extensively analyzed. The simplest Leslie model is linear in $%
{\bf n}$, so that the evolution equation is just ${\bf n}(t+1)={\Bbb A}\,%
{\bf n}(t)$. Here, ${\Bbb A}$ is the Leslie matrix 
\begin{equation}
{\Bbb A}=\left( 
\begin{array}{ccccc}
f_0 & f_1 & \ldots &  & f_D \\ 
v_0 & 0 & \ldots &  & 0 \\ 
0 & v_1 &  &  & 0 \\ 
\vdots &  & \ddots &  & \vdots \\ 
0 & \ldots & 0 & v_{D-1} & 0
\end{array}
\right) ,
\end{equation}
where the elements $f_a$ are the fecundities (number of offspring produced)
of individuals of age $a$, and the $v_a$ are Verhulst factors (the fraction
of individuals of age $a$ who survive to become age $a+1$). Though the
original Leslie model had all the $\{f_a\}$ and $\{v_a\}$ as constants,
generalizations have since been made to ${\bf n}$--dependent factors \cite
{pielou,liu,wikan,nisbet}, so that the evolution dynamics becomes inherently 
{\em non--linear}. For example, ${\bf n}$--dependent Verhulst factors are
often used to mimic competition for finite resources.

However one deficiency of the models discussed hitherto is that they are
deterministic. Real population systems are of course affected by random
fluctuations, coming from the environment and/or from the intrinsic dynamics
of the birth/death processes. Such stochastic Leslie models have also been
investigated \cite{pollard,tulj,engen}, however only for cases where the
birth/death probabilities were {\em independent} of the population vector $%
{\bf n}$. To date no information has been available regarding the more
realistic case of {\em fluctuations} in stochastic age structured models
with {\em population dependent birth/death probabilities} \cite{engen}. This
is the situation we will study in this letter.

Population models have also been intensively studied by physicists in recent
years in the context of so--called ``bit--string'' models of evolution \cite
{penna}. These models are based on the mutation accumulation hypothesis \cite
{rose}, which assumes that during the aging process each individual
accumulates exclusively late--acting deleterious genetic mutations. In
``bit--string'' models the genome of a particular species is encoded as a
series of `$0$'s and `$1$'s (deleterious mutations), and as an individual
ages the bits (genes) are activated one by one. When the accumulated sum of
bad genes reaches a certain number the individual dies (although death may
also occur at younger ages due to Verhulst competition). Note that the
``bit--strings'' of the offspring may differ from those of the parent due to
additional beneficial (`$1$'$\to$`$0$') or deleterious (`$0$'$\to$`$1$')
mutations. This type of model is clearly well suited to efficient computer
simulation \cite{penna,tol}. In its simplest case, where individuals die
after the first deleterious mutation, ``bit--string'' models simply
correspond to multiple genome population models with age structure, where
the different genomes can be distinguished by different maximum ages.
Deterministic versions of some of these models have already been treated
analytically \cite{tol}. However, an analysis of the important role played
by fluctuations has so far been lacking. Our calculations form the first
step towards filling this gap.

We begin our analysis by defining our discrete time population model. For
simplicity, we consider only a single species reproducing {\em asexually},
without mutations. Thus, our system can be described by a single vector $%
{\bf n}$. At each time step we compute the Verhulst factor $V({\bf n})$ and
let each individual survive with probability $V$. After this ``pruning'',
each of the remaining individuals of age $a$ may give birth to $F_a$
offspring with probability $r_a$. At this point the remaining population is
aged by one time step, with the exception of the new offspring who make up $%
n_0$. Individuals who exceed the maximum age $D$ then die immediately and
are removed from the system. Since our model allows for a variable
reproduction probability $r_a$ as a function of age, features such as
puberty and menopause can be naturally incorporated. However, we do assume
that reproductive individuals of the same age produce an identical number ($%
F_a$) of offsprings. Note that we are not restricting ourselves to specific
forms for $V$, $r_a$, or $F_a$ beyond some general features, so that this
analysis may easily be applied to the various ``bit--string'' models \cite
{penna,tol}. These general features include $V,r_a\in \left[ 0,1\right] $,
since they represent {\em probabilities}. Furthermore, we assume that $V$
depends only on the {\em total }population $N\equiv \sum_an_a$, via the
ratio $N/N_0$, where $N_0$ represents a characteristic population size that
the resources can support. Also, to be reasonable, $V$ is assumed to be
monotonically decreasing with $N$. For comparisons with simulations, we use $%
V=s_0\exp (-N/N_0)$, where $s_0$ is another constant, a form frequently
chosen in the biology literature \cite{murray}. In contrast, the algebraic
form $V=1-(N/N_0)$ is the favorite in the recent physics literature\cite
{penna,tol}. We prefer the exponential form, since absolute cut-offs seem
unrealistic in a real population system.

As mentioned above, {\em deterministic} versions of this model have been
studied \cite{liu,wikan,nisbet}; in particular the population dynamics of a
semivoltine species studied in Ref.~\cite{nisbet} is quite similar to a $D=1$
version of our model. Our goal is to go beyond these deterministic
treatments and analyze the fluctuations and correlations in this system.
Therefore, we need to consider $P({\bf n},t)$, the probability of finding
the population with a particular distribution ${\bf n}$ at time $t$. Its
evolution obeys the master equation 
\begin{eqnarray}
&&\qquad \qquad P({\bf n},t+1)=\sum_{m_{0},\ldots ,m_{D},n_{D+1}}P({\bf m}%
,t)\times   \nonumber \\
&&\left[ \prod_{a=1}^{D+1}\hspace{-0.1cm}%
{m_{a-1} \choose n_{a}}%
V^{n_{a}}\left[ 1-V\right] ^{m_{a-1}-n_{a}}\hspace{-0.05cm}\right] \hspace{%
-0.2cm}\sum_{\;b_{0},\ldots ,b_{D}}\hspace{-0.3cm}\delta \hspace{-0.05cm}%
\left[ n_{0}-\hspace{-0.05cm}\sum_{c=0}^{D}F_{c}b_{c}\right]   \nonumber \\
&&\qquad \qquad \times \left[ \prod_{a=1}^{D+1}%
{n_{a} \choose b_{a-1}}%
r_{a-1}^{b_{a-1}}\left[ 1-r_{a-1}\right] ^{n_{a}-b_{a-1}}\right] .
\label{mastereq}
\end{eqnarray}
Note that the $n_{D+1}$ is just a ``temporary'' variable, which keeps track
of the number of $m_{D}$'s who survive the Verhulst ``pruning'' so that they
can give birth before dying from old age. Let us emphasize that this
equation is actually quite complex, since $V$ is a function of the total
population. Multiplying (\ref{mastereq}) by $n_{c}$ and summing over all the
other indices, we obtain 
\begin{eqnarray}
&&\left\langle n_{a}\right\rangle _{t+1}=\left\langle V(N)\
n_{a-1}\right\rangle _{t},\qquad\qquad (a>0),  \label{Exact-c} \\
&&\langle n_{0}\rangle _{t+1}=\sum_{d}F_{d}r_{d}\left\langle V(N)\
n_{d}\right\rangle _{t},  \label{Exact-0}
\end{eqnarray}
where $\left\langle \bullet \right\rangle _{t}$ denotes the average of $%
\bullet $ over $P({\bf n},t)$. These equations are {\em exact}. However, due
to the presence of $N$ through $V$, all moments of $P$ may be coupled
together. The mean
field (MF) approximation consists of replacing the higher order moments by
appropriate products of the first moment. Hence we find 
\begin{eqnarray}
&&\left\langle n_{a}\right\rangle _{t+1}^{{\rm MF}}=\left[
V(\sum_{c}\left\langle n_{c}\right\rangle _{t}^{{\rm MF}})\right]
\left\langle n_{a-1}\right\rangle _{t}^{{\rm MF}},\quad (a>0),  \label{MFAc}
\\
&&\langle n_{0}\rangle _{t+1}^{{\rm MF}}=\left[ V(\sum_{c}\left\langle
n_{c}\right\rangle _{t}^{{\rm MF}})\right] \sum_{d}F_{d}r_{d}\left\langle
n_{d}\right\rangle _{t}^{{\rm MF}},  \label{MFA0}
\end{eqnarray}
where, to be clear, we have written the explicit expression for $N$. These
non-linear equations are known to contain a rich variety of behavior,
depending on the details of $F_{c},r_{c},$ and $V$. For example, if $%
\sum_{c}F_{c}r_{c}<1$, the reproductive rates are too low and the population
will eventually die out. On the other hand, if the reproductive rates are
large enough, the population will display period doubling bifurcations and
chaos \cite{guck}. Let us focus on the ``moderate'' range, so that a {\em %
simple} non--zero steady--state exists. In that case these equations are
easily solved to give 
\begin{equation}
\langle n_{a}\rangle ^{{\rm MF}}=N(z)\frac{z^{a}(1-z)}{(1-z^{D+1})},
\label{mfsolna}
\end{equation}
where $z$ is the unique, positive, real root of the equation $%
\sum_{c}F_{c}r_{c}z^{c+1}=1$, and $N(z)$ is the steady state total
population, given by the value that satisfies $V(N)=z$. Our objective is to
go beyond such well--known mean field solutions and investigate fluctuations
and correlations, i.e., the second moments of $P$. Thus, we multiply Eq.~(%
\ref{mastereq}) by $n_{a}n_{b}$ and sum over all the other indices. For $a>0$%
, $b>0$, we have 
\begin{eqnarray}
&&\left\langle n_{a}n_{b}\right\rangle _{t+1}=\left\langle
V^{2}n_{a-1}n_{b-1}\right\rangle _{t}+\delta _{ab}\left\langle
V(1-V)n_{a-1}\right\rangle _{t},\hspace{-0.1cm}  \label{exactnanb} \\
&&\left\langle n_{a}n_{0}\right\rangle _{t+1}=\sum_{c}F_{c}r_{c}\left\langle
V^{2}n_{a-1}n_{c}\right\rangle _{t}+  \nonumber \\
&&\qquad \qquad \qquad +F_{a-1}r_{a-1}\left\langle
V(1-V)n_{a-1}\right\rangle _{t},  \label{exactnan0} \\
&&\left\langle n_{0}n_{0}\right\rangle
_{t+1}=\sum_{c,d}F_{c}r_{c}F_{d}r_{d}\left\langle
V^{2}n_{c}n_{d}\right\rangle _{t}+  \nonumber \\
&&\qquad \qquad \qquad +\sum_{c}\left[ F_{c}^{2}(r_{c}\left\langle
Vn_{c}\right\rangle _{t}-r_{c}^{2}\left\langle V^{2}n_{c}\right\rangle
_{t})\right] .  \label{exactn0n0}
\end{eqnarray}
Note that, like Eqs.~(\ref{Exact-c}) and~(\ref{Exact-0}), these are {\em %
exact}. Assuming $N_{0}\gg 1$, and that the system is well away from
``critical'' points (e.g., bifurcations and the survival/extinction
transition), it is reasonable to postulate a Gaussian distribution for $%
P^{*}\left( {\bf n}\right) $ with width of $O(\sqrt{N_{0}})$. Rewriting Eqs.
(\ref{Exact-c},\ref{Exact-0}) and~(\ref{exactnanb}-\ref{exactn0n0}) for $%
\left\langle \bullet \right\rangle ^{*}$, we then have a closed set of
equations for the first and second moments. This approach should form the
first step in a systematic expansion of all quantities in decreasing powers
of $N_{0}$. Furthermore, in the same spirit, we will let ${\bf n}/N_{0}$
assume {\em continuous} values. As a check, we will compare the results from
this approach with those from a Monte--Carlo simulation, for a simple case.

Proceeding, let us write 
\begin{eqnarray}
&&P^{*}\left( {\bf n}\right) =\left( \frac{1}{2\pi N_{0}}\right) ^{(D+1)/2}%
\frac{1}{\sqrt{\det {\Bbb G}}}\times   \label{probdist} \\
&&\qquad \qquad \times \exp \left[ -\frac{1}{2N_{0}}\sum_{c,d}\left( n_{c}-%
\bar{n}_{c}\right) G_{cd}^{-1}\left( n_{d}-\bar{n}_{d}\right) \right] , 
\nonumber
\end{eqnarray}
where we expect the unknown (to be determined) parameters ${\bf \bar{n}}$
and ${\Bbb G}$ to be of $O(N_{0})$ and $O(1)$, respectively. Note that we
will integrate ${\bf n}$ from $-\infty \to \infty $ rather than from $0\to
\infty $, a simplification which will introduce only negligible errors of $%
O(\exp [-N_{0}])$. Averages can now be computed using 
\begin{equation}
\left\langle f(n_{a})\right\rangle ^{*}=f(\bar{n}_{a})+\frac{1}{2}\sum_{c,d}%
\frac{\partial ^{2}f}{\partial \bar{n}_{c}\partial \bar{n}_{d}}%
N_{0}G_{cd}+\ldots ~.  \label{funcaverage}
\end{equation}
Note that the second term in Eq.~(\ref{funcaverage}) is expected to be
suppressed by a factor of $O(1/N_{0})$ compared to the first term $f(\bar{n}%
_{a})$. Hence the right hand side of Eq.~(\ref{funcaverage}) is actually an
expansion in powers of $1/N_{0}$. This ordering allows us to set up a
systematic perturbation theory, which can be pushed to higher orders if
desired.

{} From now on we drop the bars for clarity ($\bar{n}_{a}\to n_{a}$).
Defining $\xi \equiv \sum_{c}n_{c}/N_{0}\text{ and }V^{\prime }\equiv
dV/d\xi $, we have $\partial V/\partial n_{c}=V^{\prime }(\xi )/N_{0}$ and $%
\partial ^{2}V/\partial n_{c}\partial n_{d}=V^{\prime \prime }(\xi
)/N_{0}^{2}$, {\em independent} of $c$ or $d$. Applying Eq.~(\ref
{funcaverage}) to Eq.~(\ref{Exact-c}) gives 
\begin{equation}
n_{a}=Vn_{a-1}+V^{\prime }\sum_{c}G_{ca-1}+\frac{n_{a-1}}{2N_{0}}V^{\prime
\prime }\sum_{c,d}G_{cd}+\ldots ~.  \label{na}
\end{equation}
An equation for $n_{0}$ can be similarly derived. With the assumptions ${\bf %
n}\sim \,O(N_{0})$ and ${\Bbb G}\sim \,O(1)$, the latter two terms in Eq.~(%
\ref{na}) represent $O(1/N_{0})$ corrections to the mean field results,
while the remaining (lowest order) pieces make up the mean field equation (%
\ref{MFAc}). Writing a perturbative expansion: $%
n_{a}=n_{a}^{(0)}+n_{a}^{(1)}+\ldots ~$(with $n_{a}^{(k)}{\Bbb \sim }%
~O(N_{0}^{1-k})$), we see that $n_{a}^{(0)}$ is given by Eq. (\ref{mfsolna}%
), while the first order result is 
\begin{eqnarray}
&&\qquad \qquad n_{a}^{(1)}=\sum_{c}\left( \left[ {\Bbb I-S}\right]
^{-1}\right) _{ac}U_{c},\quad {\rm with}  \label{n1} \\
&&S_{ab}\hspace{-0.06cm}=\hspace{-0.06cm}{\frac{\partial [Vn_{a-1}]}{%
\partial n_{b}}},\;\;U_{a}\hspace{-0.06cm}=\hspace{-0.06cm}{\frac{1}{2}}%
\sum_{c,d}{\frac{\partial ^{2}[Vn_{a-1}]}{\partial n_{c}\partial n_{d}}}%
N_{0}G_{cd},\;\;(a>0),  \nonumber \\
&&S_{0b}=\sum_{c}F_{c}r_{c}{\frac{\partial [Vn_{c}]}{\partial n_{b}}}%
,\;\;U_{0}={\frac{1}{2}}\sum_{c,d,e}F_{c}r_{c}{\frac{\partial ^{2}[Vn_{c}]}{%
\partial n_{d}\partial n_{e}}}N_{0}G_{de},  \nonumber
\end{eqnarray}
where ${\Bbb I}$ is the unit matrix and ${\Bbb S}$ is the stability matrix
associated with the mean field (zeroth order) stationary solution. Note that
both ${\Bbb S}$ and $U$ need to be evaluated at zeroth order only. With our
assumptions about $F_{a},r_{a}$, and $V(N)$, the eigenvalues of the
stability matrix ${\Bbb S}$ usually lie within the unit circle, implying
that our mean field solution is stable. However, for sufficiently high
reproductive rates, perturbations with $\delta n_{a}\propto n_{a}$ (i.e.,
populations with the same relative age distribution, but with with different
total numbers of individuals) can have eigenvalues of less than $-1$. This
is the signal of a period doubling bifurcation, leading to the breakdown of
our Gaussian perturbation expansion (see also below).

Applying the same analysis to the second moments, we obtain, after some
lengthy algebra, 
\begin{equation}
G_{ab}-\sum_{c,d}S_{ac}S_{bd}G_{cd}=K_{ab}\quad ,\quad \forall a,b,
\label{gab}
\end{equation}
where 
\begin{eqnarray}
&&K_{ab}=\delta _{ab}(1-V)\frac{n_a}{N_0},\qquad \qquad \qquad \quad \;a,b>0,
\label{kab} \\
&&K_{a0}=K_{0a}=F_{a-1}r_{a-1}(1-V)\frac{n_a}{N_0},\quad \;a>0,  \label{kao}
\\
&&K_{00}={\frac 1{N_0}}\sum_bVF_b^2r_bn_b(1-Vr_b).  \label{k00}
\end{eqnarray}
Again, all quantities need to be evaluated only at the zeroth order, so
that, e.g., $V$ is just $z$. In compact form this equation can be written as 
${\Bbb G}-{\Bbb S}{\Bbb G}\,{\Bbb S}^T={\Bbb K}$, which may be solved by
series 
\begin{equation}
{\Bbb G}={\Bbb K}+{\Bbb S}{\Bbb K}\,{\Bbb S}^T+{\Bbb S}^2\,{\Bbb K}\left( 
{\Bbb S}^T\right) ^2+\ldots =\sum_n{\Bbb S}^n\,{\Bbb K}\left( {\Bbb S}%
^T\right) ^n.  \label{gsersol}
\end{equation}
Since the eigenvalues and eigenvectors of ${\Bbb S}$ are known, let us write 
${\Bbb S}={\Bbb M}\,{\Bbb E}\,{\Bbb M}^{-1}$, where ${\Bbb E}$ is in Jordan
form, with the eigenvalues on the diagonal, and ${\Bbb M}$ is the matrix
(with its {\em columns}) composed of the corresponding {\em right}
eigenvectors. Note that ${\Bbb M}$ is not necessarily orthogonal or unitary.
If we define $\tilde{{\Bbb G}}={\Bbb M}^{-1}{\Bbb G}\left( {\Bbb M}%
\,^T\right) ^{-1}$ and $\tilde{{\Bbb K}}={\Bbb M}^{-1}{\Bbb K}\left( {\Bbb M}%
^{-1}\right) ^T$, then it is straightforward to show that $\tilde{{\Bbb G}}%
=\sum_n{\Bbb E}^n\tilde{{\Bbb K}}\,{\Bbb E}^n$. For simplicity, let us focus
on the case where ${\Bbb E}$ is diagonal. Then the sum is easily performed,
so that 
\begin{equation}
\tilde{G}_{ab}=\tilde{K}_{ab}/\left( 1-e_ae_b\right) \quad \quad \text{(no
sum)},  \label{final}
\end{equation}
where the $\{e\}$ are the eigenvalues. Since ${\Bbb G}={\Bbb M}\,\tilde{%
{\Bbb G}}{\Bbb M}^T$, we can directly obtain the matrix ${\Bbb G}$ and with
it all the information about the Gaussian probability distribution (\ref
{probdist}). The explicit formula for computing ${\Bbb G}$ is our principal
result. Given a particular form of $V(N)$ and reproductive parameters $%
r_a,F_a$, we can compute ${\Bbb G}$ and find the fluctuations in, as well as
the correlations between, the populations of various ages.

The result (\ref{final}) contains a further appealing feature: the signal of
bifurcation. {} From stability analysis, we know that period doubling
emerges when the eigenvalue associated with $\delta n_a\propto n_a$ reaches $%
-1$. Examining Eq.~(\ref{final}), we see that it is precisely this feature
which signals the breakdown of the Gaussian approximation. Furthermore, in
many studies of, e.g., the Penna ``bit--string'' model \cite{penna},
menopause sets in before death, so that ${\Bbb E}$ is not diagonal. Then the
final expression for $\tilde{{\Bbb G}}$ will be slightly more complicated,
although the above conclusions will remain qualitatively unchanged.

To check the above analysis, we study the simplest possible case: a $2$ age
system (i.e. $D=1$), with $r_a=F_a=1$. Choosing the exponential form for $V$
with $s_0=1$ and $N_0=100$, mean field theory yields $n_0^{(0)}=157.4$ and $%
n_1^{(0)}=119.3$. Performing our analysis, we arrive at the first order
corrections to $n_0$ and $n_1$, the fluctuations in the populations of each
age, and the correlation between the populations of the two ages. The
results are listed in Table~\ref{table1}, alongside those from Monte--Carlo
simulations \cite{robert}. The agreement is excellent, validating our
approach. Note that the corrections $n_a^{(1)}/n_a^{(0)}$ are less than 1\%,
vindicating our assertion that they should be $O(1/N_0)$.

Up to this point we have been considering models with {\em discrete} time
steps. However it is perfectly possible, and sometimes more appropriate
biologically, to analyze models in {\em continuous} time \cite
{murray,charles}. Let us conclude with some brief remarks about fluctuations
in this context. A suitable equation for the mean field population dynamics
is 
\begin{equation}
{\frac{\partial n(x,t)}{\partial t}}=-{\frac{\partial n(x,t)}{\partial x}}%
-\lambda n(x,t)\int_0^Dn(x^{\prime },t)dx^{\prime },  \label{contmfeq}
\end{equation}
with boundary conditions for birth at $x=0$ and certain death at $x=D$.
However the birth/death/aging processes giving rise to Eq.~(\ref{contmfeq})
can also be written as a ballistic reaction model on a discrete spatial
lattice but with continuous time. As is well known \cite{masfth}, starting
from the corresponding microscopic lattice master equation, techniques now
exist to map this model onto a field theory in continuous space--time. The
ensuing action can be recast as a Langevin equation, with the result being
Eq. (\ref{contmfeq}), but with extra {\em multiplicative noise} terms. The
form of these noise terms would then be completely specified, without any 
{\em ad--hoc} guesses. Unfortunately the field--theoretic action is rather
awkward, due to the presence of non--local interactions and non--local,
multiplicative noise (from fluctuations in the birth process at $x=0$).
However, simplifications occur if we are interested only in the simple,
non--zero steady state, where expansions about the mean field solution
should be adequate. In this case the leading noise terms enter {\em %
additively}, so that a perturbation theory analogous to the above approach
can be set up.

In conclusion, we have developed analytic techniques for dealing with
fluctuation effects in a general class of population models with age
structure. The results we have presented also form a first step towards an
improved analytic understanding of the ``bit--string'' models of evolution.
Finally, it would be interesting to perform further investigations near the
bifurcation point, since interesting collective behavior can be expected in
that region.

We thank R. Desai, B. Schmittmann, and U.C. T\"{a}uber for illuminating
discussions, and also R. J. Astalos for sharing simulation results prior to
publication. This research has been supported in part by the NSERC of Canada
and the US National Science Foundation through the Division of Materials
Research. 
\begin{table}[tbp]
\caption{Comparison of results for the $2$ age model.}
\label{table1}
\begin{tabular}{|c||c|c|}
\hline
& Gaussian Approximation & Simulations \\ \hline\hline
$\langle n_0\rangle$ & 156.8 & 156.4 \\ 
$\langle n_1\rangle$ & 118.9 & 118.6 \\ \hline
$\langle n_0^2\rangle-\langle n_0\rangle^2$ & 122.0 & 130.7 \\ 
$\langle n_0n_1\rangle-\langle n_0\rangle\langle n_1\rangle$ & 63.9 & 67.8
\\ 
$\langle n_1^2\rangle-\langle n_1\rangle^2$ & 69.9 & 71.7 \\ \hline
\end{tabular}
\vspace{-.2in}
\end{table}

\end{document}